\documentclass{kluwer}    

\newdisplay{guess}{Conjecture}

\begin{document}                         
\begin{article}
\begin{opening}         
\title{
Spectra of thermally unstable slim discs} 
\author{Ewa \surname{Szuszkiewicz}} 
\institute{Institute of Physics, University of Szczecin,
Szczecin, Poland \\
Toru\'n Centre for Astronomy, Nicolaus Copernicus University,
Toru\'n, Poland \\
Dipartimento di Fisica ``Galileo Galilei'', Universit\`a
degli Studi di Padova, Italy \\
International School for Advanced Studies, SISSA, Italy}
\author{Roberto \surname{Turolla}} 
\institute{Dipartimento di Fisica ``Galileo Galilei'', Universit\`a
degli Studi di Padova, Italy}
\author{Luca \surname{Zampieri}}  
\institute{Dipartimento di Fisica ``Galileo Galilei'', Universit\`a
degli Studi di Padova, Italy}
\runningauthor{Ewa Szuszkiewicz}
\runningtitle{Spectra from thermally unstable disc}
\date{}

\begin{abstract}
          Thermal instability driven by radiation pressure might be relevant
          for intrinsically bright accreting sources.  The most promising
          candidate where this instability seems to be at work is one of the
          two known galactic superluminal sources, GRS 1915+105 (Belloni
          et al. 1997). In spite of
          being of relevance, this scenario has not yet been confirmed by
          proper time-dependent modelling.
          Non-linear time-dependent calculations performed by Szuszkiewicz
          and Miller (1998) show that thermally unstable discs undergo
          limit-cycle behaviour with successive evacuation and refilling
          of the central parts of the disc. This evolution is very similar
          to the one proposed by Belloni et al. (1997) in their
          phenomenological model.
	  Further investigations are needed to confirm the thermal instability 
	  being operational in this source.
          First of all the spectra emitted from the disc during its evolution 
	  should be calculated and compared  with observations. 
	  Here such spectra are computed assuming 
	  local blackbody emission from the best studied transonic disc model. 
\end{abstract}
\keywords{accretion discs, thermal instability, spectra}
\end{opening}           
\section{Limit-cycle behaviour of accretion flows}  
The simplest disc model which properly describes physical conditions around 
black holes is a transonic (slim) accretion disc. It is an axisymmetric,
vertically integrated, non-self-gravitating, differentially rotating fluid
configuration.  The assumption about
self-gravity being negligible defines the outer edge of the disc
and the requirement that the radial velocity equals the local sound speed 
somewhere in the disc, and then exceeds it, determines its radial
structure. 
Once the mass, $M$, of the central black hole,
accretion rate, $\dot M$, and viscosity coefficient $\alpha$ are
specified the structure of the stationary disc is uniquely determined.  
However, such a disc might be subject to secular, thermal and 
pulsational instabilities.
Here we focus on the thermal instability in which a temperature perturbation 
out of the equilibrium is enhanced by thermal processes. Moreover, we 
consider the thermal instability which is triggered  by one particular 
mechanism, namely  by the radiation pressure.
The possibility that a global instability might develop in this model
was indicated by the results of a local stability analysis. 
In the temperature - surface density diagram
the sequence of equilibrium models forms 
a characteristic S-shape curve. Such curves reflect the  possibility of
a transition between low and high accretion rate states, associated
with an increase of the temperature.
\begin{figure}[h] 
\vspace{7.5cm}
\includegraphics{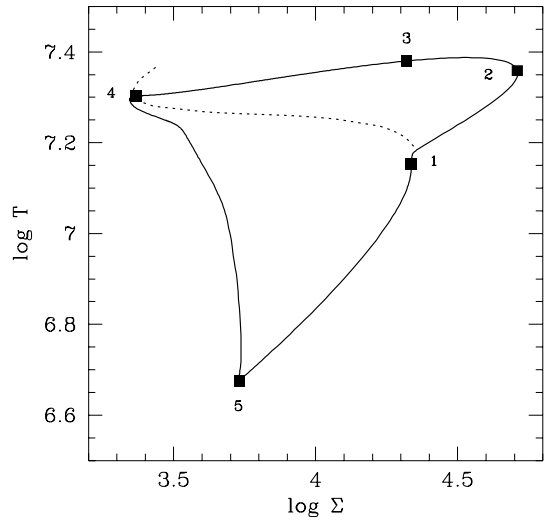}
\caption[]{The phase portrait in the $\log T$-$\log\Sigma$ 
plane for $r=10r_G$ (here $r_G$ is the Schwarzschild radius).
 The dotted line represents the equilibrium states, the so called
S-curve. The lower branch of the S-curve overlaps
with the trajectory of the evolving model. The continuous line shows the
trajectory of the model during the evolution. The points marked 
1 to 5 indicate those stages for which spectra are
calculated. The points 1 to 4 correspond to the phase of disc evacuation,
the point 5 is at the beginning of the refilling phase.}
\label{dot}
\end{figure}
The first complete calculation of the global evolution of thermal instability
in transonic discs was performed by Szuszkiewicz and Miller (1998).
They have shown that at least for some values of $M$, $\dot M$ and $\alpha$
this evolution has a  cyclic character and consists of successive
evacuation and refilling phases.
An example of the limit-cycle
for the best studied thermally unstable model with $M=10M_{\odot}$, 
$\dot M = 0.06$ (in Eddington units) and $\alpha =0.1$ is shown
in Figure 1. 
This particular non-stationary disc behaviour
might have interesting observational consequences.

\section{Spectra in the blackbody approximation}

The global dynamical calculations, described in Section 1,
determine the overall properties of the transonic
disc evolution. However, in order to get detailed theoretical predictions
for the directly observed quantities, we must compute
the spectra emitted from thermally unstable
discs during the evacuation and refilling of their central parts. 
\begin{figure}[h] 
\vspace{7.5cm}
\includegraphics{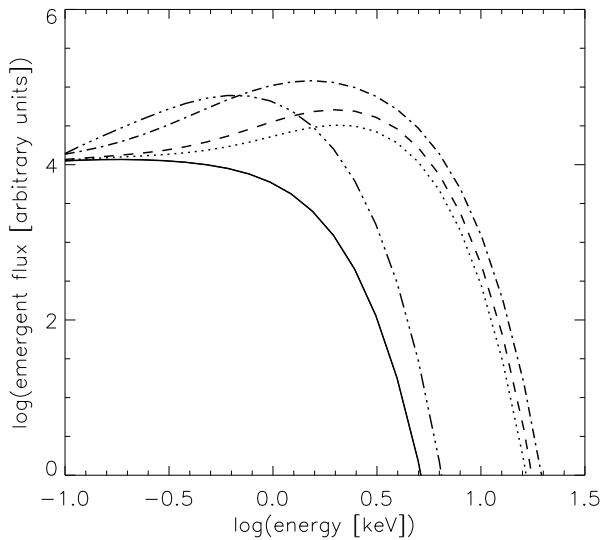}
\caption[]{The disc spectra for the evolutionary stages 
shown in Fig. 1: stage 1 (solid line), stage 2 (dotted line),
stage 3 (dashed line), stage 4 (dash-dotted line), and stage 5 (triple 
dashed-dotted line).}
\label{spec}
\end{figure}
In the case of a geometrically thin accretion disc, it is possible
to consider the vertical and radial disc structures separately.
It is also reasonable to assume local heat balance. In order to 
calculate the spectrum, it is sufficient to divide the disc into
a number of concentric annuli, calculate the spectrum
emitted by each annulus and then sum them all together.
The simplest assumption is that each annulus radiates like a blackbody at 
the local effective temperature.
We stress that this assumption does not necessarily apply to
all relevant physical
situations. Most of the previous studies were concentrated on
the
modelling of the vertical structure and radiative transfer 
through each annulus in Shakura-Sunyaev discs (e.g. Shimura and Takahara, 
1995).
However, the thin disc hypothesis breaks down at high accretion
luminosities and it may not be a good approximation at very low accretion
rates either. 
Slim disc models are more appropriate for astrophysical applications
(Szuszkiewicz et al., 1996; Wang et al., 1999).
Here as a first approximation we have calculated spectra assuming that 
the disc radiates locally as a blackbody (Figure 2). 
During the disc evacuation phase the spectra become significantly harder.
Detailed spectra calculations, which account properly for radiative
transfer in the disc atmospheric layers are under way and will be
presented in the forthcoming paper 
(Szuszkiewicz, Turolla and Zampieri, in preparation).

\section{Concluding remarks}

Transonic discs with accretion rates
larger than a critical value which depends on the black hole mass
and the viscosity parameter, show time
dependent, cyclic behaviour due to the onset of thermal instabilities
driven by
radiation pressure. This behaviour can be described in terms of the 
cycle period, the duration of the bursting and quiescent phases, 
the amplitude and shape
of the burst and finally by the spectral variability. 
For the different parameters of the system ($M$, $\alpha$, $\dot M$)
we expect different variability properties and emitted spectra.
Detailed theoretical predictions 
are definitely needed for modelling high-resolution timing and
spectral data presently available for black hole systems.

\acknowledgements
E.S. gratefully acknowledges financial support from the Polish State Committee
for Scientific Research (grant KBN 2P03D01817),
the Italian MURST and the University of Padova.
Work partially supported through MURST grant COFIN9802154100.



\end{article}
\end{document}